\documentclass[%
 reprint,
 amsmath,amssymb,
 aps,
 pra,
]{revtex4-1}

\usepackage{graphicx}
\usepackage{dcolumn}
\usepackage{lipsum}
\usepackage{bm}
\usepackage{hyperref}

\begin{document}

\preprint{APS/123-QED}

\title{Laser-cooled caesium atoms confined with magic-wavelength dipole inside a hollow-core photonic-bandgap fiber}

\author{Taehyun Yoon}
\email{th2yoon@uwaterloo.ca}
\author{Michal Bajcsy}%
\affiliation{%
Institute for Quantum Computing, University of Waterloo, 200 University Ave W, Waterloo, ON N2L 3G1, Canada\\
Department of Electrical and Computer Engineering, University of Waterloo, 200 University Ave W, Waterloo, ON N2L 3G1, Canada
}%

\date{\today}

\begin{abstract}
We report loading of laser-cooled caesium atoms into a hollow-core photonic-bandgap fiber and confining the atoms in the fiber's 7 $\mu$m diameter core with a magic-wavelength dipole trap at $\sim$935 nm. The use of the magic wavelength removes the AC-Stark shift of the 852nm optical transition in caesium caused by the dipole trap in the fiber core and suppresses the inhomogeneous broadening of the atomic ensemble that arises from the radial distribution of the atoms. This opens the possibility to continuously probe the atoms over time scales of a millisecond -- approximately 1000-times longer than what was reported in previous works, as dipole trap does not have to be modulated. We describe our atom loading setup and its unique features and present spectroscopy measurements of the caesium's D$_2$ line in the continuous wave dipole trap  with up to $1.7 \times 10^{4}$ loaded inside the hollow-core fiber.
\end{abstract}

\maketitle

Cold atoms confined inside hollow-core waveguides offer a unique platform for studies of light-matter interactions, quantum and non-linear optics, and effective photon-photon interactions mediated by atomic ensembles\cite{Chang2014, Carusotto2013}. Over the past decade, several groups reported loading of cold atomic ensembles into the core of microstructured hollow-core fibers, starting with the transfer of a Bose-Einstein condensate from a free-space into a fiber-guided dipole trap \cite{Christensen2008}. This was followed by laser-cooled rubidium experiments \cite{Bajcsy2011, Blatt2014, Langbecker2017, Xin2018, Hilton2018}, which demonstrated electromagnetically induced transparency (EIT) and all-optical switching \cite{Bajcsy2009}, optical memory and stationary light pulses \cite{Blatt2016}, Rydberg excitations \cite{Langbecker2017}, and atom interferometry \cite{Xin2018} in this platform. At the same time, proposals to use on-chip hollow-core waveguides instead of fibers have been put forward \cite{Bappi2017, Bitarafan2017} although the on-chip waveguides for now carry the penalty of significantly higher propagation losses compared to the fibers.

In addition to the above-listed experiments performed with cold atoms in hollow-core fibers, similar non-linear and quantum optics experiments at low light levels have also been realized with room-temperature atoms loaded inside these fibers \cite{Ghosh2006, Light2007a, Saha2011, Venkataraman2011, Venkataraman2012, Perrella2013, Perrella2013a, Sprague2014, Epple2014}. However, in these room temperature systems, the thermal motion of the atoms limits the effective atom-photon interaction probability because of the inhomogeneous Doppler broadening of the optical transitions, and the coherence times are limited by the atoms' collisions with the walls of the waveguide core. The latter can be somewhat mitigated by using fibers with a large diameter hollow core, although at the expense of a further decrease of interaction probability between a single atom and a single photon. 
Laser cooled atoms offer the advantages of negligible Doppler broadening and the suppression of atom-wall collision by confining the atoms inside the waveguide with an optical dipole trap. Unfortunately, the AC-Stark shift the dipole trap light induces in the confined atoms also changes the central frequency of the optical transitions adds an inhomogeneous broadening, both of which depend on the power of the dipole trap and the radial distribution of the atoms in the fiber \cite{Bajcsy2011}. To avoid this, the intensity of the dipole trap is modulated at a rate of $\sim$1 MHz \cite{Bajcsy2009, Blatt2014, Blatt2016, Xin2018, Hilton2018}. The atoms are probed when the trapping light is off, and the trap is then turned back on  to recapture the atoms before they collide with the walls and are lost. This approach removes the AC-Stark shift effects but the time window during which the atoms can be probed in the absence of inhomogeneous broadening is now limited by the transverse temperature of the atoms and the diameter of the fiber core. The probing window can be potentially extended by transverse cooling of the atoms, which was proposed and observed in Ref. \cite{Peyronel2012}, or, if the experiment does not demand maximizing of the interaction probability between a single atom and a single photon, by selecting a fiber with a large diameter hollow core.   

\begin{figure*}
\includegraphics[width=1.0\textwidth]{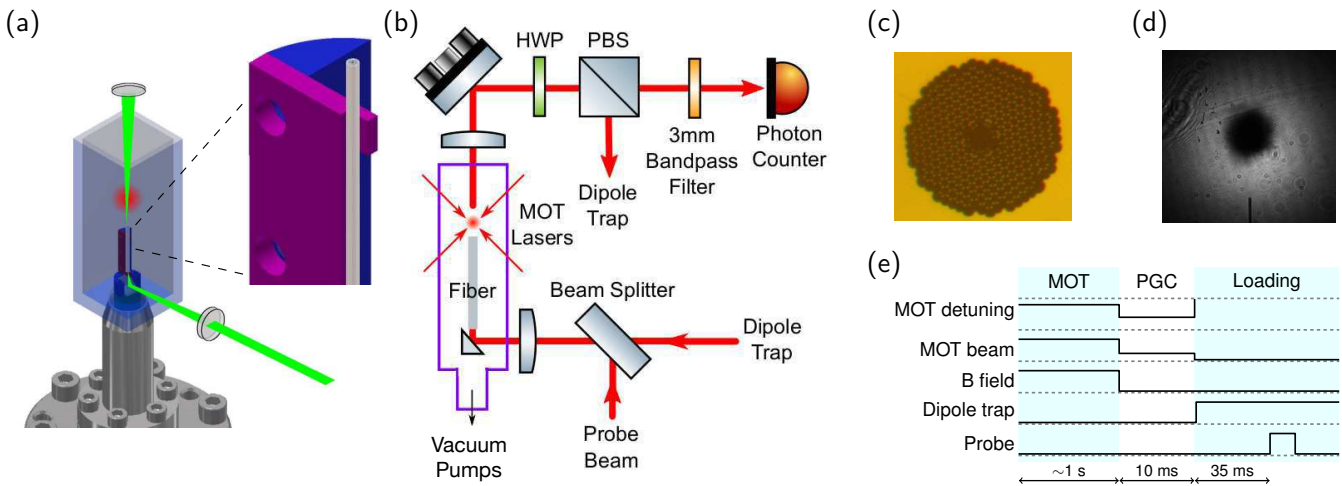}
    \caption{(a) The glass cell containing the MOT and the hollow-core fiber and a detail of the fiber mounting structure.  (b) Schematic of the overall experimental setup. (c) Microscope image of the cross section of the hollow-core fiber. (c) Absorption image of the atomic cloud right after polarization gradient cooling. The fiber tip cab be seen at the bottom. (d) Time sequence of our experimental procedure.}
\label{fig:figure1}
\end{figure*}

An alternative approach was recently demonstrated by Katori group \cite{Okaba2014} who used a ``magic wavelength'' dipole trap \cite{Katori1999} to avoid the undesired effects of the dipole trap by choosing a wavelength for the dipole trap that shifts both the ground and the excited state of the target optical transition in the same direction and by the same amount. In their experiment, Okaba \textit{et al.} loaded laser-cooled strontium atoms into an optical lattice at 914 nm formed by two counter-propagating laser beams inside a kagome lattice fiber with a $\sim$40 $\mu$m diameter hypocycloid hollow core \cite{Bradley2013} and were able to observe an optical transition with a linewidth of 7.8 kHz.
We utilize this ``magic wavelength'' dipole trap approach in the work presented here as well but with the ultimate goal of engineering low-power optical nonlinearities, which are enhanced  by tight confinement of photons. Such tight confinement can be provided by hollow core fibers that confine light through the photonic-bandgap effect \cite{Cregan1999} and can guide light in modes with diameter of just a few micrometers. Caesium atoms are then a convenient choice for this type of experiments as the D$_2$ line optical transition at 852 nm has a magic wavelength at $\sim$935 nm \cite{McKeever2003} and both wavelengths are guided by a commercially available photonic-bandgap fibre with a 7 $\mu$m diameter core.
Here, we report loading and confining of $\sim 1.7 \times 10^{4}$ laser-cooled caesium atoms into a such fiber. We describe our experimental setup and present the spectroscopy results observed in the presence of the 935 nm trapping light.   

Fig.~\ref{fig:figure1}(a) and (b) shows our experimental setup. A 20mm-long piece of a hollow-core fiber (HC-800-02 from NKT Photonics) with fundamental mode diameter of $\sim$5.5 $\mu$m (Fig.~\ref{fig:figure1}(c)) is vertically mounted in the lower half of a ColdQuanta glass cell with anti-reflection coatings on both the inside and outside walls. The fiber piece is glued with a low-outgassing epoxy onto a blade-shaped mount. This mount is designed to minimize the obstruction of the magneto-optical trap (MOT) beams and to allow optical access to the fiber from the side for optical pumping and imaging in the future. An non-evaporable getter pump and a small ion pump maintain the pressure inside the cell at an ultra-high vacuum level ($\sim 10^{-10}$ torr). The whole vacuum system is mounted on a pair of 1m-long CNC rails and can be moved out of the optics setup without disrupting it. This is intended to allow changing of the fiber piece in the future with minimized experimental downtime. 
The dipole trap at 935 nm is provided by a Ti:Sapph laser (SolsTiS by M Squared). A perpendicularly polarized weak probe beam is combined to the dipole beam using a 90:10 beam splitter. An aspheric lens ($f$ = 18 mm) located outside the vacuum cell couples the combined beams into the fiber at an efficiency of 40$\sim$50 \%. Another aspheric lens ($f$ = 50 mm) collimates the light coming out the fiber, and the dipole beam is filtered out using a combination of a half-wave plate (HWP) and polarizing beam splitter (PBS), as well as a 3 nm-wide bandpass filter centered at 852 nm. Finally, a single photon counting module (Excelitas Technologies Corp, SPCM-AQRH-NIR) detects the signal transmitted through a single-mode fiber patch cable (Fig.~\ref{fig:figure1}(b)). Note that another 3 nm bandpass filter centered at 935 nm cleans the light from the Ti:Sapph laser before it is sent into the experiment to remove the spontaneous emission coming from the gain medium of this laser. While the Ti:Sapph cavity suppresses this spontaneous emission by $\sim$100 dB to a level where it does not affect the atoms in the fiber, the emission shows up as a background noise on the single photon detectors in the absence of this filter when the atoms are probed in a continuous wave dipole trap. 

The caesium D2 transition line (6$^{2}$S$_{1/2} \rightarrow \ ^{2}$P$_{3/2}$, 852 nm) is used in conducting our experiment. We cool and trap $\sim10^{8}$ caesium atoms in a MOT $\sim$5 mm above the tip of the fiber (Fig.~\ref{fig:figure1}(d)). The trapping beams are detuned by $3 \times \Gamma / 2\pi$ from the cycling transition $\vert F = 4 \rangle \rightarrow \vert F'=5 \rangle$. The repumper beam, resonant on the transition $\vert F = 3 \rangle \rightarrow \vert F'=3 \rangle$, recycles atoms in the dark state $\vert F = 3 \rangle$ back into the cycling transition.

We then further cool the trapped atoms using polarization gradient cooling (PGC) for 10 ms \cite{Dalibard1989}. The lower temperature of atoms enables the more atoms to be loaded inside the fiber more efficiently. While we have managed to cool our atomic cloud down to 12 $\mu$K, the experiments were mostly performed at a temperature of $\sim$30 $\mu$K as an interplay between additional factors besides the atoms' temperature -- such as the atomic cloud size, shape, and the position -- determines the number of atoms loading into the fiber. Once this additional cooling process is finished, the trapping and repump beams are shut off by acousto-optic modulator (AOM) switches and the atoms start to free-fall because of gravity. At the same time, the dipole trap laser turns on, and the dipole optical potential generated by the diverging dipole laser guides the atoms into the core of the fiber where the light intensity is high. Figure.~\ref{fig:figure1}(e) depicts the control sequence of the experiment. 

\begin{figure}
\includegraphics[width=1.0\columnwidth]{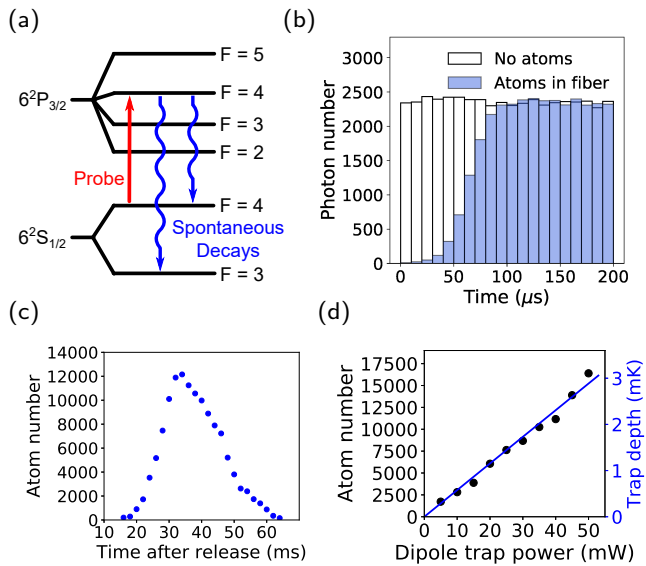}
\caption{(a) Diagram of the hyperfine energy levels involved in the "bleaching" measurement of the atom number. (b) Histogram of photon counts with and without atoms inside the fiber showing counts accumulated from 30 cooling and loading cycles. (c) Measured atom numbers with the cooling laser shut off at 0 ms and 40 mW (measured right above the cell) of 935 nm fiber-coupled dipole beam on continuously. (d) Measured atom numbers (black circles) and calculated optical dipole trap depth (blue line) versus the dipole trap power.}
\label{fig:figure2}
\end{figure}

To determine the number of atoms loaded into the fiber, we measure the time dependent transmission of a calibrated probe through the system \cite{Blatt2014}.
The probe ($\sim$ 10 pW) is tuned to an open transition $\vert F = 4 \rangle \rightarrow \vert F' = 4  \rangle$, and we tag the arrival time of transmitted photons. In a baseline measurement, when atoms are not loaded into the fiber, the probe photons transmission through the system is uniform as a function of time other than the small fluctuations in the photon numbers (white histogram in Fig.~\ref{fig:figure2}(a)). When the caesium atoms are loaded inside the fiber, the probe photons are initially fully absorbed by the atoms, which get excited into the $\vert F' = 4 \rangle$ state. As the excited atoms decay to the $\vert F = 3 \rangle$, which is a dark state for the probe beam, the medium becomes transparent (``bleached'') as time goes on (blue histogram in Fig.~\ref{fig:figure2}(b)). The difference in the transmitted photon numbers with and without loaded atoms indicates how many photons are absorbed by atoms. Taking into account the branching ratio from state $\left | F' = 4 \right \rangle$ (7/12 to state $\left | F = 4 \right \rangle$) and the efficiency of the photon counter, we derive the number of caesium atoms inside the fiber. Note that the result of this measurement is affected neither by the radial distribution of the atoms inside the fiber, nor by the probe being not precisely on resonance with the atomic transition, as these conditions will only change the rise time in the probe pulse transmission but not the total number of absorbed photons. As a result, the measurement can be done with both modulated and continuous wave dipole trap. Lastly, since atoms that are outside the fiber will see a significantly lower probe intensity and will thus take a much longer time than the atoms inside the fiber to be pumped into the dark state. This effect then limits the error in the estimate of the number of atoms loaded inside the fiber that arises from the unloaded atoms -- an error that can be significant in a measurement of  transmission as a function of frequency. 
Fig.~\ref{fig:figure2}(c) then shows how the atom number inside fiber evolves after the release of the atomic cloud from the cooling region. At time 0 ms, the trapping and cooling lasers are shut off, and the atomic cloud starts to fall. The atoms begin to appear inside the fiber around 20 ms after release. As more atoms arrive, the atomic population rapidly increases and peaks at 33 ms, then mostly goes away before 65 ms. This time scale is consistent with atomic motion due to gravity -- an initially stationary atom 5 mm above the fiber tip arrives at the top end of the fiber at 29 ms and reaches the bottom end at 48 ms due to the acceleration from the $\sim$2 mK dipole trap. We observed a linear scaling of the maximum number of atoms detected inside the fiber with the dipole trap power measured at the top side of the hollow-core fiber piece (Fig.~\ref{fig:figure2}(d)).  Overall, with a dipole trap power of 50 mW, we loaded $1.7 \times 10^{4}$ atoms into the fiber core from $1.05 \times 10^{8}$ atoms in the MOT, corresponding to the loading efficiency of $\sim 1.6 \times 10^{-4}$.

\begin{figure}
\includegraphics[width=1.0\columnwidth]{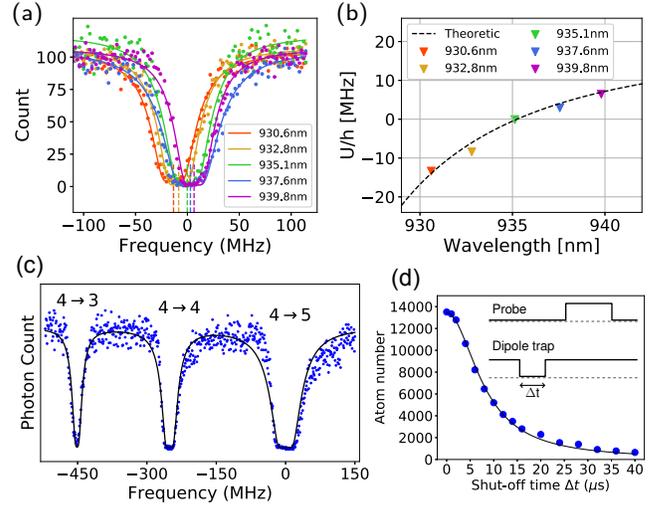}
\caption{\label{fig:figure3} (a) Transmission of weak probe light as a function of its frequency for different dipole trap wavelengths. 0 MHz marks the resonant frequency of the $\vert F=4 \rangle \rightarrow \vert F'=5 \rangle$ transition. (b) Resonant transition frequencies extracted from (a) as a function of the dipole trap wavelength compared to a theoretical prediction (black dashed line). (c) Observed transmission (blue circles) of the probe frequency scanned over transitions to different hyperfine levels in the excited state $5^{2}P_{3/2}$ from the ground $\vert F=4\rangle$ state with a fit of Eq.~(\ref{eq:long_scan}) (black solid line). (d)The time sequence for the temperature measurement inside the fiber (inset) and the atom numbers in the recaptured cloud inside the fiber as a function of the dipole trap off time. A fit based on the model Eq.~(\ref{eq:recapture}), shown in black line, gives a transverse temperature of $\sim$ 2.3mK.} 
\end{figure}

Figure~\ref{fig:figure3}(a) presents our measurements of the transmission of a weak probe ($\sim$ 1 pW) through the fiber as a function of the probe's detuning from the closed transition ($\vert F = 4 \rangle \rightarrow \vert F' = 5 \rangle$) for different wavelengths of the continuous dipole trap. At $\lambda_{dipole}$=935.1 nm, the center of the absorption is at the resonance frequency of this transition. The solid lines correspond to fitting the experimental data  $T=P/P_{0}$, where $P_{0}$ and $P$ are the power of the probe transmitted through the fiber without and with the atoms loaded into the fiber, with a Lorentzian dependence on probe detuning 
\begin{equation}
T(\omega) =  \textrm{exp} \left [ -\frac{\mathcal{D}_{opt}} { (1 + 4 ((\omega - \omega_{0}) / \gamma)^{2})} \right ].
\end{equation}
Here, $\mathcal{D}_{opt}$ is the resonant optical depth of the system,  $\omega$ is the frequency of the probe light, $\omega_{0}$ is the center frequency, and $\gamma$ is the coherence decay. The solid line fits in Fig.~\ref{fig:figure3}(a) were obtained with $\mathcal{D}_{opt}$ and $\omega_{0}$ as fitting parameters and $\gamma$ set to 5.2MHz, corresponding to the decay rate of the $5^{2}P_{3/2}$ state. Setting $\gamma$ to a different value had little effect on the value of the center frequency $\omega_{0}$ obtained from the fits, but the value of 5.2MHz resulted in the best fits of the experimental data. We were thus able to cancel the overall frequency shift of the ensemble by selecting a particular wavelength of the dipole trap, as well as to suppress the inhomogeneous broadening arising from the AC Stark shifts varying with atoms' radial position in the fiber. In general though, the value of $\gamma$ that resulted in best fit varied within a factor of two in our experiments, depending on the number of atoms loaded and the transition probed. To suppress the inhomogeneous broadening reliably , we would need to better control the polarization of the dipole trap inside the fiber and prepare the atoms in a single Zeeman sublevel state \cite{Okaba2014, Liu2017}.

Figure~\ref{fig:figure3}(b) compares the central frequencies to the theoretical prediction based on AC-Stark shifts at the effective dipole trap intensity of $I = \mu P / (\pi \sigma_{0}^{2})$ \cite{LeKien2013}. Here, $P$ = 40 mW is the dipole trap power in this measurement, and the $\sigma_{0}$ = 2.75 $\mu$m is the mode radius of the hollow-core fiber. The factor $\mu$ is determined by the radial distribution of the atomic ensemble inside the fiber and has its biggest value, $\mu$ = 2, when all the atoms are aligned along the fiber axis. By fitting to these data, we obtain the distribution factor $\mu$ = 1.61 of our system.

Fig.~\ref{fig:figure3}(c) then presents a transmission profile of a the probe beam scanned over the allowed transitions from the ground state $\vert F = 4 \rangle$ in the presence of continuous dipole trap beam $\lambda_{dipole}$= 935 nm. The black solid line shows the fitting curve
\begin{equation}
T(\omega) =  \sum_{j=3}^{5}\textrm{exp} \left [ -\frac{\mathcal{D}_{opt}^{(4\rightarrow j)}} { (1 + 4 ((\omega - \omega_{0}^{(4j)}) / \gamma)^{2})} \right ].
\label{eq:long_scan}
\end{equation}
where the resonant optical depths of the individual transitions, $\mathcal{D}_{opt}^{(4\rightarrow j)}$, were the only fitting parameters and the central frequencies of each absorption dip $\omega_{0}^{(4j)}$ were pre-determined to the theoretical values of the caesium hyperfine structure\cite{SteckCs}. 

It was, however, recently reported that micro lensing effect of the hollow core fiber may lead to a significant over-estimation of the optical depth in the transmission profile \cite{Noaman2018}, and this effect may also explain the fitted curve not quite following the data points in the wings of the absorption profile of the $\vert F = 4 \rangle \rightarrow \vert F'=5 \rangle$ transition in Fig.~\ref{fig:figure3}(c). 
By comparing the results of the atom counting and the optical depth measurements, we found that on the $\vert F = 3 \rangle \rightarrow \vert F'=2 \rangle$ transition, $\sim$360 atoms were needed to create a medium with $\mathcal{D}_{opt}$=1. For this measurement, the system had non-zero transmission on resonance, so the resonant optical depth could be determined directly without any assumptions about the inhomogeneous broadening. Taking into account the relative strengths of the $\vert F = 3 \rangle \rightarrow \vert F'=2 \rangle$ and the $\vert F = 4 \rangle \rightarrow \vert F'=5 \rangle$ transitions, $\sim$210 atoms are needed to create a medium with $\mathcal{D}_{opt}$=1 on the $\vert F = 4 \rangle \rightarrow \vert F'=5 \rangle$ transition. Hence, the average probability of a single photon interacting with a single atom on the strongest transition of the $D_2$ line in our system is $\sim$0.5 $\%$. 

Lastly, we characterize the transverse temperature of the atomic cloud inside the fiber with a time-of-flight measurement. Here, the dipole trap is turned off for a short time and the atomic cloud is allowed to expand, which leads to loss of atoms as they collide with the wall of the fiber core. We then turn on the dipole trap again to recapture the remaining atoms and count them with the "bleaching'' procedure from Fig.~\ref{fig:figure2}. The time sequence for this process is shown in Fig.~\ref{fig:figure3}(d) together with the number of recaptured atoms at various shut-off times of the dipole trap. The number of recaptured atoms at time $\tau_{r}$ follows the relation of
\begin{equation}
N(\tau_{r}) \approx N_{0} \left ( 1 - \text{exp} \left [ \dfrac{-(R_{\text{core}}/r_{0})^{2}}{1 + (v_{0}/r_{0})^2 \tau_{r}^{2}} \right ] \right ),
\label{eq:recapture}
\end{equation}
where $v_{0} = \sqrt{(kT/m_{Cs})}$ is the most probable velocity and $R_{\text{core}}$ is the fiber hollow-core radius \cite{Bajcsy2011}. By fitting this equation to the data set, we estimate the transverse temperature of the atomic cloud to be $\sim$2.3 mK for a 935 nm dipole trap power of 40 mW. This is very close to the calculated optical potential depth 2.31 mK at this dipole trap power and wavelength. The atoms thus heat up significantly as they are loaded into the fiber-guided dipole trap. Note that since the "bleaching'' procedure gives a more reliable atom number count that is unaffected by the atoms' radial distribution, we should obtain a more precise temperature estimate than the original approach from Ref. \cite{Bajcsy2011}, in which the number of remaining atoms was estimated from an optical depth measurement.
Using a transverse cooling method, such as the one outlined in Ref. \cite{Peyronel2012}, it should be possible to decrease this rather high transverse temperature, which should increase the lifetime of the atoms inside the fiber, as well as the average probability of single photon interacting with a single atom in this system.

To summarize, we demonstrated the loading of up to $1.7 \times 10^{4}$ caesium atoms into a photonic-bandgap fiber with a 7 $\mu$m-diameter hollow core using a magic wavelength dipole trap, which allows us to probe the atoms without having to modulate the dipole trap and opens possibilities to realize experiments that can't be performed in the short time windows of the modulated dipole trap approach. This is the first time cold atom trapping with a magic wavelength dipole trap has been demonstrated in a hollow-core fiber with a small enough core so that the single photon interaction probability with a single atom, which scales as an inverse of the mode area, is $\sim$0.5 $\%$.
Our system offers an excellent platform to study tantalizing topics in quantum nonlinear optics, such as strongly interacting photons \cite{Chang2008, Hafezi2012}, superradiance \cite{Goban2015}, and dynamical control of photonic bandgap\cite{Andre2002}, and can potentially be used to realize a steady-state fiber-integrated super-radiant laser \cite{Bohnet2012}. Additional areas of enhanced light-matter interaction can be accessed by integrating cavities into the fiber, such as proposed in Ref. \cite{Flannery2017} or demonstrated in Ref. \cite{Flannery2018}.

\begin{acknowledgments}
This work was supported by Industry Canada, NSERC's Discovery grant, and by Ontario's Ministry of Innovation Early Researcher Award.
We wish to thank Dr. C. Haapamaki for his help with assembling the vacuum system and the AOM controllers.
\end{acknowledgments}

\bibliography{cs_in_magic}

\end{document}